\documentclass[aps,reprint]{revtex4}
\usepackage{amssymb}
\usepackage{amsmath}

\input{tcilatex}
\begin{document}

\title{Collective modes, AC response and magnetic properties of the 3D Dirac
semi-metal in the triplet superconducting state. }
\author{B. Rosenstein$^{1}$, B.Ya. Shapiro$^{2}$ and I. Shapiro$^{2}$}
\affiliation{$^{1}$Department of Electrohysics, National Chiao Tung University, Hsinchu,
Taiwan, R.O.C. and Ariel University, Israel. }
\affiliation{$^{2}$Department of Physics, Institute of Superconductivity, Bar-Ilan
University, 52900 Ramat-Gan, Israel.}

\begin{abstract}
It was recently shown that conventional phonon-electron interactions may
induce a triplet pairing state in time-reversal invariant 3D Dirac semi -
metals. Starting from the microscopic model of the isotropic Dirac
semi-metal, the Ginzburg-Landau equations for the vector order parameter is
derived using the Gor'kov technique. The collective modes including gapless
Goldstone modes, and gapped Higgs modes of various polarizations are
identified. They are somewhat analogous to the modes in the B phase of $%
He^{3}$, although in the present case quantitatively there is a pronouneced
difference between longitudinal and transverse components. The difference is
caused by the vector nature of the order parameter leading to two different
coherence lengths or penetration depths. The system is predicted to be
highly dissipative due to the Goldstone modes. The time dependent Ginzburg -
Landau model in the presence of external fields is used to investigate some
optical and magnetic properties of such superconductors. The AC conductivity
of a clean sample depends on the orientation of the order parameter. It is
demonstrated that the difference between the penetration depths results in
rotation of the polarization vector of microwave passing a slab made of this
material. The upper critical magnetic field $H_{c2}$ was found. It turns out
that at fields close to $H_{c2}$ the order parameter orients itself
perpendicular to the field direction. In certain range of parameters the
triplet superconducting phase persists at arbitrarily high magnetic field
like in some $p$ wave superconductors.
\end{abstract}

\maketitle

\section{ Introduction}

Recently 3D Dirac semi - metals (DSM) like $Na_{3}Bi$ and $Cd_{3}As_{2}$
with electronic states described by Bloch wave functions, obeying the
"pseudo-relativistic" Dirac equation (with the Fermi velocity $v_{F}$
replacing the velocity of light) were observed\cite{Potemski} and attracted
widespread attention. The discovery of the 3D Dirac materials makes it
possible to study their physics including remarkable electronic properties.
This is rich in new phenomena like giant diamagnetism that diverges
logarithmically when the chemical potential approaches the 3D Dirac point, a
linear-in-frequency AC conductivity that has an imaginary part\cite{Wan},
quantum magnetoresistance showing linear field dependence in the bulk\cite%
{Ogata}. Most of the properties of these new materials were measured at
relatively high temperatures. However recent experiments at low temperature
on topological insulators and suspected 3D Dirac semi-metals exhibit
superconductivity. Early attempts to either induce or discover
superconductivity in Dirac materials were promising. The well known
topological insulator $Bi_{2}Se_{3}$ doped with $Cu$, becomes
superconducting at $T_{c}=3.8K$\cite{Ong}. At present its pairing symmetry
is unknown. Some experimental evidence\cite{phononexp} point to a
conventional phononic pairing mechanism. The spin independent part of the
effective electron - electron interaction due to phonons was studied
theoretically\cite{phonontheory}.\ For a conventional parabolic dispersion
relation, typically independent of spin, the phonon mechanism leads to the $%
s $-wave superconductivity. The layered, non-centrosymmetric heavy element $%
PbTaSe_{2}$ was found to be superconducting \cite{Ong}. Its electronic
properties like specific heat, electrical resistivity, and
magnetic-susceptibility indicate that $PbTaSe_{2}$ is a moderately coupled,
type-II BCS superconductor with large electron-phonon coupling constant of $%
\lambda =0.74$. It was shown theoretically to possess a very asymmetric 3D
Dirac point created by strong spin-orbit coupling. If the 3D is confirmed,
it might indicate that the superconductivity is conventional phonon mediated.

More recently when the $Cu$ doped $Bi_{2}Se_{3}$ was subjected to pressure%
\cite{pressureBiSe}, $T_{c}$ increased to $7K$ at $30GPa$. Quasilinear
temperature dependence of the upper critical field $H_{c2},$ exceeding the
orbital and Pauli limits for the singlet pairing, points to the triplet
superconductivity. The band structure of the superconducting compounds is
apparently not very different from its parent compound $Bi_{2}Se_{3}$, so
that one can keep the two band $\mathbf{k}\cdot \mathbf{p}$ description ($Se$
$p_{z}$ orbitals on the top and bottom layer of the unit cell mixed with its
neighboring $Bi$ $p_{z}$ orbital). Electronic-structure calculations and
experiments on the compounds under pressure\cite{pressureBiSe} reveal a
single bulk three-dimensional Dirac cone like in $Bi$ with large spin-orbit
coupling. Moreover very recently some pnictides were identified as
exhibiting Dirac spectrum. This effort recently culminated in discovery of
superconductivity in $Cd_{3}As_{2}$\cite{3Dsuper}. It is claimed that the
superconductivity is $p$-wave at least on the surface.

The case of the Dirac semi-metals is very special due to the strong spin
dependence of the itinerant electrons' effective Hamiltonian. It was pointed
out\cite{Fu} that in this case the triplet possibility can arise although
the triplet gap is smaller than that of the singlet, the difference
sometimes is not large for spin independent electron - electron
interactions. Very recently the spin dependent part of the phonon induced
electron - electron interaction was considered\cite{DasSarma14} and it was
shown that the singlet is still favored over the triplet pairing. Another
essential spin dependent effective electron-electron interaction is the
Stoner exchange among itinerant electrons leading to ferromagnetism in
transition metals. While in the best 3D Weyl semi-metal candidates it is too
small to form a ferromagnetic state, it might be important to determine the
nature of the superconducting condensate. It turns out that it favors the
triplet pairing\cite{Rosenstein15}. Also a modest concentration of magnetic
impurities makes the triplet ground state stable.

In a multicomponent superconductor collective modes including gapless
Goldstone modes (sound), and gapped Higgs modes of various polarizations
play an important role in determining thermal and optical properties of the
material\cite{Brusov}. In addition, as mentioned above, generally the
applied magnetic field is an ultimate technique to probe the superconducting
state. In a growing number of experiments, in addition to magnetotransport,
magnetization curves, the magnetic penetration depth and upper critical
magnetic field were measured\cite{magnetization}.{\huge \ }It is therefore
of importance to construct a Ginzburg - Landau (GL) description\cite%
{Ketterson} of these novel materials. This allows to study inhomogeneous
order parameter configurations (junctions, boundaries, etc.), the collective
modes (somewhat analogous to the modes in the B phase of $He^{3}$ and
magnetic and optical response that typically involve inhomogeneous
configurations (like vortices) not amenable to a microscopic description.

In the present paper we derive such a GL type theory for triplet
superconductor from the microscopic isotropic DSM model with attractive
local interaction. The order parameter in this case is a vector field and
the GL theory of vector field already considered in literature\cite%
{Knigavko,Machida,Babaev} in connection with putative $p$ - wave
superconductors have several extraordinary features, both quantitative and
qualitative.

The paper is organized as follows. The model of the (phonon mediated or
unconventional) local interactions of 3D Dirac fermion is presented and the
method of its solution (in the Gorkov equations form) including the symmetry
analysis of possible pairing channels and the vectorial nature of the
triplet order parameter is given in Section II. In Section III the Gorkov
formalism, sufficiently general to derive the GL equations, is briefly
presented. The most general form of the GL energy of the triplet
superconductor in magnetic field consistent with the symmetries is given in
IV. The coefficient of the relevant terms are calculated from the
microscopic DSM model in section V. Section VI is devoted to applications of
the GL model. The ground \ state degeneracy, the character of its
excitations and basic magnetic properties are discussed. The vector order
parameter is akin to optical phonons with sharp distinction between
transverse and longitudinal modes. Transverse and longitudinal coherence
lengths and penetration depths are calculated and the upper critical
magnetic field is discussed. Section VI includes generalizations to include
Pauli paramagnetism, discussion of an experimental possibility of
observation of the excitation and\ conclusion.

\section{The local pairing model in the Dirac semi-metal.}

\subsection{Pairing Hamiltonian in the Dirac semi-metal.}

Electrons in the 3D Dirac semimetal are described by field operators $\psi
_{fs}\left( \mathbf{r}\right) $, where $f=L,R$ are the valley index
(pseudospin) for the left/right chirality bands with spin projections taking
the values $s=\uparrow ,\downarrow $ with respect to, for example,\ the $z$
axis. To use the Dirac ("pseudo-relativistic") notations, these are combined
into a four component bi-spinor creation operator, $\psi ^{\dagger }=\left(
\psi _{L\uparrow }^{\dagger },\psi _{L\downarrow }^{\dagger },\psi
_{R\uparrow }^{\dagger },\psi _{R\downarrow }^{\dagger }\right) $, whose
index $\gamma =\left\{ f,s\right\} $ takes four values. The non-interacting
massless Hamiltonian with Fermi velocity $v_{F}$ and chemical potential $\mu 
$ reads\cite{Wang13}%
\begin{eqnarray}
K &=&\int_{\mathbf{r}}\psi ^{+}\left( \mathbf{r}\right) \widehat{K}\psi
\left( \mathbf{r}\right) \text{;\ \ }  \label{kinetic} \\
\text{\ \ \ \ }\widehat{K}_{\gamma \delta } &=&-i\hbar v_{F}\nabla
^{i}\alpha _{\gamma \delta }^{i}-\mu \delta _{\gamma \delta }\text{,}
\end{eqnarray}%
where the three $4\times 4$ matrices, $i=x,y,z$, 
\begin{equation}
\mathbf{\alpha }=\left( 
\begin{array}{cc}
\mathbf{\sigma } & 0 \\ 
0 & -\mathbf{\sigma }%
\end{array}%
\right) \text{,}  \label{alpha}
\end{equation}%
are presented in the block form via Pauli matrices $\mathbf{\sigma }$. They
are related to the Dirac $\mathbf{\gamma }$ matrices (in the chiral
representation, sometimes termed "spinor") by $\ \mathbf{\alpha }=\beta 
\mathbf{\gamma }$ with%
\begin{equation}
\beta =\left( 
\begin{array}{cc}
0 & \mathbf{1} \\ 
\mathbf{1} & 0%
\end{array}%
\right) \text{.}  \label{beta}
\end{equation}%
Here $\mathbf{1}$ is the $2\times 2$ identity matrix.

We consider a special case of 3D rotational symmetry that in particular has
an isotropic Fermi velocity. Moreover we assume time reversal, $\Theta \psi
\left( \mathbf{r}\right) =i\sigma _{y}\psi ^{\ast }\left( \mathbf{r}\right) $%
, and inversion symmetries although the pseudo- Lorentz symmetry will be
explicitly broken by interactions. The spectrum of single particle
excitations is linear. The chemical potential $\mu $ is counted from the
Dirac point.

As usual in certain cases the actual interaction can be approximated by a
model local one:

\begin{equation}
V_{eff}=\mathbf{-}\frac{g}{2}\int_{\mathbf{r}}\psi _{\alpha }^{+}\left( 
\mathbf{r}\right) \psi _{\beta }^{+}\left( \mathbf{r}\right) \psi _{\beta
}\left( \mathbf{r}\right) \psi _{\alpha }\left( \mathbf{r}\right) \text{.}
\label{local}
\end{equation}%
Unlike the free Hamiltonian $K$, Eq.(\ref{kinetic}), this interaction
Hamiltonian does not mix different spin components.

Spin density in Dirac semi-metal has the form 
\begin{equation}
\mathbf{S}\left( \mathbf{r}\right) =\frac{1}{2}\psi ^{+}\left( \mathbf{r}%
\right) \mathbf{\Sigma }\psi \left( \mathbf{r}\right) \text{,}
\label{spindensity}
\end{equation}%
where the matrices 
\begin{eqnarray}
\mathbf{\Sigma }\mathbf{=-\alpha } &&\gamma _{5}=\left( 
\begin{array}{cc}
\mathbf{\sigma } & 0 \\ 
0 & \mathbf{\sigma }%
\end{array}%
\right) ,  \label{sigma_matrices} \\
\gamma _{5} &=&\left( 
\begin{array}{cc}
-\mathbf{1} & 0 \\ 
0 & \mathbf{1}%
\end{array}%
\right) ,  \notag
\end{eqnarray}%
are also the rotation generators.

\subsection{The symmetry classification of possible pairing channels.}

Since we consider the local interactions as dominant, the superconducting
condensate (the off-diagonal order parameter) will be local%
\begin{equation}
O=\int_{\mathbf{r}}\psi _{\alpha }^{+}\left( \mathbf{r}\right) M_{\alpha
\beta }\psi _{\beta }^{+}\left( \mathbf{r}\right) ,  \label{O}
\end{equation}%
where the constant matrix $M$ should be a $4\times 4$ antisymmetric matrix.
Due to the rotation symmetry they transform covariantly under infinitesimal
rotations generated by the spin $S^{i}$ operator, Eq.(\ref{spindensity}):

\begin{eqnarray}
&&\int_{\mathbf{r,r}^{\prime }}\left[ \psi _{\alpha }^{+}\left( r\right)
M_{\alpha \beta }\psi _{\beta }^{+}\left( r\right) ,\psi _{\gamma
}^{+}\left( r^{\prime }\right) \mathbf{\Sigma }_{\gamma \delta }^{i}\psi
_{\delta }\left( r^{\prime }\right) \right]  \label{transformation} \\
&=&-\int_{r}\psi _{\gamma }^{+}\left( r\right) \left( \mathbf{\Sigma }%
_{\gamma \delta }^{i}M_{\delta \kappa }+M_{\gamma \delta }\mathbf{\Sigma }%
_{\delta \gamma }^{ti}\right) \psi _{\kappa }^{+}\left( r\right) \text{.} 
\notag
\end{eqnarray}%
Here and in what follows "$t$" denotes the transpose matrix. The
representations of the rotation group therefore characterize various
possible superconducting phases.

Out of 16 matrices of the four dimensional Clifford algebra six are
antisymmetric and one finds one vector and three scalar multiplets of the
rotation group. The multiplets contain:

(i) a triplet of order parameters: 
\begin{eqnarray}
&&\left\{ M_{x}^{T},M_{y}^{T},M_{z}^{T}\right\}  \label{triplet} \\
&=&\left\{ -\beta \alpha _{z},-i\beta \gamma _{5},\beta \alpha _{x}\right\}
=\left\{ T_{x},T_{y},T_{z}\right\}  \notag
\end{eqnarray}%
The algebra is 
\begin{equation}
\mathbf{\Sigma }_{i}T_{j}+T_{j}\mathbf{\Sigma }_{i}^{t}=2i\varepsilon
_{ijk}T_{k}\text{.}  \label{algebra}
\end{equation}%
Note that the three matrices $T_{i}$ are Hermitian.

(ii) three singlets

\begin{equation}
M_{1}^{S}=i\alpha _{y};\text{ \ \ }M_{2}^{S}=i\Sigma _{y};\text{ \ \ }%
M_{3}^{S}=-i\beta \alpha _{y}\gamma _{5}\text{.}  \label{singlets}
\end{equation}%
Which one of the condensates is realized at zero temperature is determined
by the parameters of the Hamiltonian and is addressed next within the
Gaussian approximation. As was shown in our previous work \cite%
{Rosenstein15,Li}, either exchange interactions or magnetic impurities make
the triplet state a leading superconducting channel in these materials.
Therefore we will consider in the next section only the vector channel.

\section{Gorkov equations and the triplet pairing}

\subsection{Gorkov equations for Green's functions in matrix form}

Using the standard BCS formalism, the Matsubara Green's functions ($\tau $
is the Matsubara time) 
\begin{eqnarray}
G_{\alpha \beta }\left( \mathbf{r},\tau ;\mathbf{r}^{\prime },\tau ^{\prime
}\right) &=&-\left\langle T_{\tau }\psi _{\alpha }\left( \mathbf{r},\tau
\right) \psi _{\beta }^{\dagger }\left( \mathbf{r}^{\prime },\tau ^{\prime
}\right) \right\rangle \text{;}  \label{GFdef} \\
F_{\alpha \beta }\left( \mathbf{r},\tau ;\mathbf{r}^{\prime },\tau ^{\prime
}\right) &=&\left\langle T_{\tau }\psi _{\alpha }\left( \mathbf{r},\tau
\right) \psi _{\beta }\left( \mathbf{r}^{\prime },\tau ^{\prime }\right)
\right\rangle \text{;}  \notag \\
F_{\alpha \beta }^{+}\left( \mathbf{r},\tau ;\mathbf{r}^{\prime },\tau
^{\prime }\right) &=&\left\langle T_{\tau }\psi _{\alpha }^{\dagger }\left( 
\mathbf{r},\tau \right) \psi _{\beta }^{\dagger }\left( \mathbf{r}^{\prime
},\tau ^{\prime }\right) \right\rangle \text{,}  \notag
\end{eqnarray}%
obey the Gor'kov equations\cite{AGD}:%
\begin{gather}
-\frac{\partial G_{\gamma \kappa }\left( \mathbf{r},\tau ;\mathbf{r}^{\prime
},\tau ^{\prime }\right) }{\partial \tau }-\int_{\mathbf{r}^{\prime \prime
}}\left\langle \mathbf{r}\left\vert \widehat{K}_{\gamma \beta }\right\vert 
\mathbf{r}^{\prime \prime }\right\rangle G_{\beta \kappa }\left( \mathbf{r}%
^{\prime \prime },\tau ;\mathbf{r}^{\prime },\tau ^{\prime }\right)
\label{Gorkov} \\
-gF_{\beta \gamma }\left( \mathbf{r},\tau ;\mathbf{r},\tau \right) F_{\beta
\kappa }^{+}\left( \mathbf{r},\tau ,\mathbf{r}^{\prime },\tau ^{\prime
}\right) =\delta ^{\gamma \kappa }\delta \left( \mathbf{r-r}^{\prime
}\right) \delta \left( \tau -\tau ^{\prime }\right) ;  \notag \\
\frac{\partial F_{\gamma \kappa }^{+}\left( \mathbf{r},\tau ;\mathbf{r}%
^{\prime },\tau ^{\prime }\right) }{\partial \tau }-\int_{\mathbf{r}^{\prime
\prime }}\left\langle \mathbf{r}\left\vert \widehat{K}_{\gamma \beta
}^{t}\right\vert \mathbf{r}^{\prime \prime }\right\rangle F_{\beta \kappa
}^{+}\left( \mathbf{r}^{\prime \prime },\tau ;\mathbf{r}^{\prime },\tau
^{\prime }\right)  \notag \\
-gF_{\gamma \beta }^{+}\left( \mathbf{r},\tau ;\mathbf{r},\tau \right)
G_{\beta \kappa }\left( \mathbf{r},\tau ,\mathbf{r}^{\prime },\tau ^{\prime
}\right) =0\text{.}  \notag
\end{gather}%
These equations are conveniently presented in matrix form (superscript $t$
denotes transposed and $I$ - the identity matrix): 
\begin{eqnarray}
&&\int_{X^{\prime \prime }}\left[ 
\begin{array}{c}
D^{-1}\left( X,X^{\prime \prime }\right) G\left( X^{\prime \prime
},X^{\prime }\right) - \\ 
-\Delta \left( X\right) F^{+}\left( X,X^{\prime }\right)%
\end{array}%
\right] =  \label{matrixeq} \\
&=&I\delta \left( X-X^{\prime }\right) \text{;}  \notag
\end{eqnarray}%
\begin{equation*}
\int_{X^{\prime \prime }}D^{t-1}\left( X,X^{\prime \prime }\right)
F^{+}\left( X^{\prime \prime },X^{\prime }\right) +\Delta ^{t\ast }\left(
X\right) G\left( X,X^{\prime }\right) =0\text{.}
\end{equation*}%
Here $X=\left( \mathbf{r},\tau \right) $, $\Delta _{\alpha \beta }\left(
X\right) =gF_{\beta \alpha }\left( X,X\right) $ and 
\begin{eqnarray}
D_{\alpha \beta }^{-1}\left( X,X^{\prime }\right) &=&-\delta _{\alpha \beta }%
\frac{\partial }{\partial \tau }\delta \left( X-X^{\prime }\right) -
\label{invD} \\
&&-\delta \left( \tau -\tau ^{\prime }\right) \left\langle \mathbf{r}%
\left\vert \widehat{K}_{\alpha \beta }\right\vert \mathbf{r}^{\prime
}\right\rangle \text{.}  \notag
\end{eqnarray}

In the homogeneous case the Gor'kov equations for Fourier components of the
Green's functions simplify considerably:

\begin{eqnarray}
D^{-1}\left( \omega ,\mathbf{p}\right) G\left( \omega ,\mathbf{p}\right)
-\Delta F^{+}\left( \omega ,\mathbf{p}\right) &=&I\text{;}
\label{Gorkov_uniform} \\
\widetilde{D}^{-1}\left( \omega ,\mathbf{p}\right) F^{+}\left( \omega ,%
\mathbf{p}\right) +\Delta ^{t\ast }G\left( \omega ,\mathbf{p}\right) &=&0%
\text{.}  \notag
\end{eqnarray}%
The matrix gap function can be chosen as 
\begin{equation}
\Delta _{\beta \gamma }=gF_{\gamma \beta }\left( 0\right) =\Delta
_{z}M_{\gamma \beta }\text{,}  \label{delta}
\end{equation}%
with real constant $\Delta _{z}$. Here $D^{-1}\left( \omega ,\mathbf{p}%
\right) =i\omega +\mu -\mathbf{\alpha \cdot p}$, is the noninteracting
inverse Dirac Green's function for the Hamiltonian Eq.(\ref{kinetic}) and $%
\widetilde{D}^{-1}\left( \omega ,\mathbf{p}\right) =i\omega -\mu -\mathbf{%
\alpha }^{t}\mathbf{\cdot p}$, where $\omega _{n}=\pi T\left( 2n+1\right) $
is the fermionic Matsubara frequency.

Solving these equations one obtains (in matrix form) 
\begin{eqnarray}
G^{-1} &=&D^{-1}+\Delta \widetilde{D}\Delta ^{t\ast }\text{;}
\label{solution} \\
F^{+} &=&-\widetilde{D}\Delta ^{t\ast }G\text{,}  \notag
\end{eqnarray}%
with the gap function to be found from the consistency condition 
\begin{equation}
\Delta ^{t\ast }=-g\sum\limits_{\omega p}\widetilde{D}\Delta ^{t\ast }G\text{%
.}  \label{gap eq}
\end{equation}%
Now we find solutions of this equation for each of the possible
superconducting phases.

\subsection{Homogeneous triplet solution of the gap equation.}

In this phase rotational symmetry is spontaneously broken simultaneously
with the electric charge $U\left( 1\right) $ (global gauge invariance)
symmetry. Assuming $z$ direction of the $p$ - wave condensate the order
parameter matrix takes a form: 
\begin{equation}
\Delta =\Delta _{z}T_{z}=\Delta _{z}\beta \alpha _{x},  \label{deltaT1}
\end{equation}%
where $\Delta _{z}$ is a constant.{\huge \ }The energy scale will be set by
the Debye cutoff $T_{D}$ of the electron - phonon interactions, see below.

The spectrum of elementary excitations at zero temperature was discussed in
ref. \cite{Rosenstein15}. There is a saddle point with energy gap $2\Delta
_{z}$ on the circle $p_{\perp }^{2}\equiv p_{x}^{2}+p_{y}^{2}=\mu
^{2}/v_{F}^{2},$ $p_{z}=0$. The gap $\Delta _{z}$ as a function of the
dimensionless phonon-electron coupling $\lambda =gN$, where $N$ being the
density of states (all spins and valleys included), increases upon reduction
in $\mu $. At large $\mu >>T_{D}$, as in BCS, the gap becomes independent of 
$\mu $ and one has the relation%
\begin{equation}
\frac{1}{g}=\frac{N}{12}\sinh ^{-1}\frac{T_{D}}{\Delta _{z}};N=\frac{2\mu
^{2}}{\pi ^{2}v_{F}^{3}\hbar ^{3}}\text{,}  \label{gapBCS_T}
\end{equation}%
leading to an exponential gap dependence on $\lambda $ when it is small: $%
\Delta _{z}=T_{D}/\sinh \left( 12/\lambda \right) \simeq
2T_{D}e^{-12/\lambda }$.

The critical temperature is obtained from Eq.(\ref{gap eq}) with disctret $%
\omega $ by substituting $\Delta _{z}=0$. To utilize the orthonormality of $%
T_{i}$, Tr$\left( T_{i}T_{j}^{\ast }\right) =4\delta _{ij}$, one multiplies
the gap equation by the matrix $T_{z}/g$ and takes the trace: 
\begin{equation}
\frac{1}{g}\text{Tr}\left( T_{z}T_{z}^{\ast }\right) =\frac{4}{g}=T_{c}%
\mathcal{B}_{zz}\text{.}  \label{Tc}
\end{equation}%
The bubble integral is 
\begin{eqnarray}
\mathcal{B}_{ij} &=&\sum\limits_{\omega \mathbf{p}}\text{Tr}\left( T_{i}%
\widetilde{D}T_{j}^{\ast }D\right) =4\delta _{ij}T_{c}\times  \label{Bij} \\
&&\times \sum\limits_{n\mathbf{p}}\frac{v_{F}^{2}\left( p_{\perp
}^{2}-p_{z}^{2}\right) +\mu ^{2}+\omega _{n}^{2}}{\omega _{n}^{4}+\left(
v_{F}^{2}p^{2}-\mu ^{2}\right) ^{2}+2\omega _{n}^{2}\left(
v_{F}^{2}p^{2}+\mu ^{2}\right) }\text{.}  \notag
\end{eqnarray}%
Performing first the sum over Matsubara frequencies and then integrate over $%
q$ one obtains, similarly to the singlet BCS, (see Appendix A for details):

\begin{equation}
T_{c}=\frac{2\gamma _{E}}{\pi }T_{D}e^{-12/\lambda }\text{,}  \label{Tcres}
\end{equation}%
where $\log \gamma _{E}=0.577$ is the Euler constant.

\section{A general GL description of a triplet superconductor in a magnetic
field.}

In this section the effective description of the superconducting condensate
in terms of the varying (on the mesoscopic scale) order complex parameter
vector field $\Delta _{i}\left( \mathbf{r}\right) $ is presented.

\subsection{The GL description for a vector order parameter}

The static phenomenological description is determined by the GL free energy
functional $F\left[ \mathbf{\Delta }\left( \mathbf{r}\right) ,\mathbf{A}%
\left( \mathbf{r}\right) \right] $ expanded to second order in gradients and
fourth order in $\Delta $. In a magnetic field $\mathbf{B}$, as usual, space
derivatives of the microscopic Hamiltonian become covariant derivatives $%
\mathbf{\nabla \rightarrow }\mathcal{D}\mathbf{=\nabla +}i\frac{e^{\ast }}{%
\hbar c}\mathbf{A}$, $e^{\ast }=2e$ due to gauge invariance under $\Delta
_{i}\rightarrow e^{i\chi \left( \mathbf{r}\right) }\Delta
_{i},A_{i}\rightarrow A_{i}-\frac{\hbar c}{e^{\ast }}\nabla \chi $. Naively
the only modification of the GL energy is in the gradient term, Eq.(\ref%
{gradterms}); the most general gradient term consistent with rotation
symmetry and the $U\left( 1\right) $ gauge symmetry is 
\begin{equation}
F_{grad}=N\int_{\mathbf{r}}\left\{ 
\begin{array}{c}
u_{T}\left\{ 
\begin{array}{c}
\left( \mathcal{D}_{j}\Delta _{i}\right) ^{\ast }\left( \mathcal{D}%
_{j}\Delta _{i}\right) - \\ 
-\left( \mathcal{D}_{i}\Delta _{j}\right) ^{\ast }\left( \mathcal{D}%
_{j}\Delta _{i}\right)%
\end{array}%
\right\} \\ 
+u_{L}\left( \mathcal{D}_{i}\Delta _{j}\right) ^{\ast }\left( \mathcal{D}%
_{j}\Delta _{i}\right)%
\end{array}%
\right\} \text{.}  \label{gradterms}
\end{equation}

The factor $N,$ the density of states, is customarily introduced into energy 
\cite{Rosenstein15}. It was noted in \cite{Machida}, that, unlike in the
usual scalar order parameter case, the longitudinal and transverse
coefficients are in general different, leading to two distinct coherence
lengths, see Section V. Possibilities for the local terms are\cite{Knigavko}%
\begin{equation}
F_{loc}=N\int_{\mathbf{r}}\left\{ \alpha \left( T-T_{c}\right) \Delta
_{i}^{\ast }\Delta _{i}+\frac{\beta _{1}}{2}\left( \Delta _{i}^{\ast }\Delta
_{i}\right) ^{2}+\frac{\beta _{2}}{2}\left\vert \Delta _{i}\Delta
_{i}\right\vert ^{2}\right\} \text{.}  \label{potterms}
\end{equation}

The magnetic part, $F_{mag}=B^{2}/8\pi $, completes the free energy.

\subsection{The set of time independent GL equations for triplet order
parameter}

The set of the GL equations corresponding to this energy are obtained by
variation with respect to $\Psi _{j}^{\ast }$ and $A_{i}$. The first is:

\begin{equation}
-\left\{ u_{T}\left( \delta _{ij}\mathcal{D}^{2}-\frac{1}{2}\left\{ \mathcal{%
D}_{i},\mathcal{D}_{j}\right\} \right) +\frac{1}{2}u_{L}\left\{ \mathcal{D}%
_{i},\mathcal{D}_{j}\right\} \right\} \Delta _{j}+  \label{GLEZ}
\end{equation}%
\begin{equation*}
+\alpha \left( T-T_{c}\right) \Delta _{i}+\beta _{1}\Delta _{i}\Delta
_{j}^{\ast }\Delta _{j}+\beta _{2}\Delta _{i}^{\ast }\Delta _{j}\Delta _{j}=0%
\text{.}
\end{equation*}%
The anticommutator appears due complex conjugate terms in Eq.(\ref{gradterms}%
)\cite{Maki}. The Maxwell equation for the supercurrent density is:

\begin{equation}
\ J_{i}=\frac{ie^{\ast }}{\hbar \ }N\left( u_{T}\Delta _{j}^{\ast }\mathcal{D%
}_{i}\Delta _{j}+u\Delta _{j}^{\ast }\mathcal{D}_{j}\Delta _{i}\right) +cc%
\text{,}  \label{current}
\end{equation}%
where $u=u_{L}-u_{T}$.

Having defined coefficients $u_{T,L}$,$\beta _{1,2}$ and $\alpha $, our aim
in the next Section is to deduce them from the microscopic Dirac semi-metal
model.

\section{GL coefficients from the Gor'kov equations}

For the calculation of coefficients of the local part, the homogeneous
Gor'kov equation, Eq.(\ref{gap eq}) suffices, while for calculation of the
gradient terms a general linearized equation, Eq.(\ref{matrixeq}) is
necessary.

\subsection{Local (potential) terms in Gor'kov}

Iterating once the equation Eq.(\ref{gap eq}) with help of Eq.(\ref{solution}%
) one obtains the local terms to third order in the gap function:

\begin{eqnarray}
&&\frac{1}{g}\Delta ^{\ast t}+  \label{expanded} \\
&&+\sum\limits_{\omega \mathbf{p}}\left\{ 
\begin{array}{c}
\widetilde{D}\left( \omega ,\mathbf{p}\right) \Delta ^{\ast t}D\left( \omega
,\mathbf{p}\right) - \\ 
-\widetilde{D}\left( \omega ,\mathbf{p}\right) \Delta ^{\ast t}D\left(
\omega ,\mathbf{p}\right) \Delta \widetilde{D}\left( \omega ,\mathbf{p}%
\right) \Delta ^{\ast t}D\left( \omega ,\mathbf{p}\right)%
\end{array}%
\right\} \text{.}  \notag
\end{eqnarray}%
Using $\Delta ^{t\ast }=\Delta _{i}^{\ast }T_{i}$, multiplying by $T_{i}^{t}$
and taking the trace, one gets the linear local terms%
\begin{equation}
N\alpha \left( T-T_{c}\right) \Delta _{i}^{\ast }=\frac{4}{g}\Delta
_{i}^{\ast }\text{,}  \label{linear1}
\end{equation}%
where the bubble integral was given in Eq.(\ref{Bij}). Expressing $g$ via $%
T_{c}$, see Eq.(\ref{Tcres}) allows to write the coefficient $a$ of $\Delta
_{i}^{\ast }$ in the Gorkov equation Eq.(\ref{gap eq}) is%
\begin{equation}
\alpha \left( T-T_{c}\right) =\frac{8\mu ^{2}}{3\pi ^{2}v_{F}^{3}\hbar ^{3}N}%
\log \frac{T}{T_{c}}\approx \frac{4}{3}\frac{T-T_{c}}{T_{c}}\text{.}
\label{a(T)}
\end{equation}%
The cubic terms in Eq.(\ref{expanded}), multiplied again by $T_{i}^{t}$ and
"traced" take the form 
\begin{eqnarray}
&&N\left( \beta _{1}\Delta _{j}^{\ast }\Delta _{j}\Delta _{i}^{\ast }+\beta
_{2}\Delta _{j}^{\ast }\Delta _{j}^{\ast }\Delta _{i}\right)  \label{terms}
\\
&=&-\Delta _{j}^{\ast }\Delta _{k}\Delta _{l}^{\ast }\sum\limits_{\omega 
\mathbf{p}}\text{Tr}\left\{ T_{i}^{t}\widetilde{D}T_{j}^{t\ast }DT_{k}%
\widetilde{D}T_{l}^{t\ast }D\right\} \text{.}  \notag
\end{eqnarray}%
The calculation is given in Appendix A and results in: 
\begin{equation}
\beta _{1}=\frac{7\zeta \left( 3\right) }{20\pi ^{2}}\frac{1}{T_{c}^{2}};%
\text{ \ \ }\beta _{2}=-\frac{1}{3}\beta _{1}\text{.}  \label{betas}
\end{equation}%
The Riemann zeta function is $\zeta \left( 3\right) =1.2$.

\subsection{Linear gradient terms}

To calculate the gradient terms, one first linearizes the Gor'kov equations,
Eq.(\ref{matrixeq})%
\begin{eqnarray}
&&\int_{X^{\prime \prime }}D^{-1}\left( X,X^{\prime \prime }\right) G\left(
X^{\prime \prime },X^{\prime }\right)  \label{Gorkovlin} \\
&=&I\delta \left( X-X^{\prime }\right) \rightarrow G=D^{-1}\text{;}  \notag
\\
&&F^{+}\left( X,X^{\prime }\right)  \notag \\
&=&-\int_{X^{\prime \prime }}D^{t}\left( X-X^{\prime \prime }\right) \Delta
^{\ast t}\left( X^{\prime \prime }\right) D\left( X^{\prime \prime
}-X^{\prime }\right) \text{.}  \notag
\end{eqnarray}%
In particular,%
\begin{eqnarray}
\frac{1}{g}\Delta ^{\ast t}\left( X\right) &=&F^{+}\left( X,X\right)
\label{DeltaX} \\
&=&-\int_{X^{\prime }}D^{t}\left( X-X^{\prime }\right) \Delta ^{\ast
t}\left( X^{\prime }\right) D\left( X^{\prime }-X\right) \text{.}  \notag
\end{eqnarray}%
The anomalous Green's functions are no longer space translation invariant,
so that the following Fourier transform is required: The (time independent)
order parameter is also represented via Fourier components $\Delta ^{\ast
}\left( X\right) =\sum_{\mathbf{P}}e^{-i\mathbf{P}\cdot \mathbf{r}}\Delta
^{\ast }\left( \mathbf{P}\right) $. The linear part Gor'kov equation (this
time including nonlocal parts) depending on the "external" momentum $\mathbf{%
P}$ reads:%
\begin{equation}
\frac{1}{g}\Delta ^{\ast t}\left( \mathbf{P}\right) +\sum_{\omega \mathbf{p}}%
\widetilde{D}\left( \omega ,\mathbf{p}\right) \Delta ^{t\ast }\left( \mathbf{%
P}\right) D\left( \omega ,\mathbf{p-P}\right) \text{.}  \label{DeltaP}
\end{equation}

To find the coefficients of the gradient terms, one should consider
contributions quadratic in $P$ from the expansion of both $\Delta ^{t\ast
}\left( \mathbf{P}\right) $ and $D\left( \omega ,\mathbf{p-P}\right) $. In
view of the gap equation Eq.(\ref{Tc},\ref{Bij}), the expansions of $\Delta
^{t\ast }\left( \mathbf{P}\right) $ cancel each other up to small
corrections of order $T-T_{c}$. So that multiplying by $\,T_{i}^{t}$ and
taking the trace

\begin{equation}
\frac{1}{2}P_{k}P_{l}\sum_{\omega \mathbf{p}}\text{Tr}\left\{ T_{i}^{t}%
\widetilde{D}\left( \omega ,\mathbf{p}\right) T_{j}^{t\ast }D_{kl}^{\prime
\prime }\left( \omega ,\mathbf{p}\right) \right\} \Delta _{j}^{\ast }\text{,}
\label{nonlocal _ terms}
\end{equation}%
where $D_{kl}^{\prime \prime }\left( \omega ,\mathbf{p}\right) \equiv \frac{%
\partial ^{2}D\left( \omega ,\mathbf{p}\right) }{\partial p_{k}\partial p_{l}%
}$. Comparing this with the gradient terms in the GL equation, Eq.(\ref{GLEZ}%
), see Appendix B for details, one deduces

\begin{equation}
u_{T}=\frac{28\zeta \left( 3\right) }{15\pi ^{2}}\frac{v_{F}^{2}\hbar ^{2}}{%
T_{c}^{2}};\text{ \ }u_{L}=\frac{1}{32}u_{T}\text{.}  \label{u_res}
\end{equation}%
Note the very small longitudinal coefficient, $u_{L}<<u_{T}$. As we shall
see in the following section it has profound phenomenological consequences.

\section{Basic properties of the triplet superconductor}

\subsection{Ground state structure and degeneracy}

A ground state is characterized by three independent parameters
corresponding to three Goldstone bosons. The GL energy is invariant under
both the vector $O(3)$ space rotations, $\Delta _{i}\rightarrow R_{ij}\Delta
_{j}$, and the superconducting phase $U(1)$, $\Delta _{i}\rightarrow
e^{i\chi }\Delta _{i}$. In the superconducting state characterized by the
vector order parameter $\mathbf{\Delta }$ ($\left\vert \mathbf{\Delta }%
\right\vert =\Delta $, energy gap) the $U\left( 1\right) $ is broken: $%
U\left( 1\right) \rightarrow 1$, while the $O(3)$ is only partially broken
down to its $O(2)$. There are therefore three Goldstone modes. Here we
explicitly parametrize these degrees of freedom by phases following ref.\cite%
{Knigavko}. Generally a complex vector field can be written as

\begin{equation}
\mathbf{\Delta }=\Delta \left( \mathbf{n}\cos \chi \mathbf{+}i\mathbf{m}\sin
\chi \right) ,  \label{parametrization}
\end{equation}%
where $\mathbf{n}$ and $\mathbf{m}$ are arbitrary unit vectors and $0<\chi
<\pi /2$, see Fig. 1.\ \bigskip

Using this parametrization the homogeneous part of the free-energy density,
Eq.(\ref{potterms}), takes the form

\begin{equation}
\frac{f_{loc}}{N}=\left\{ 
\begin{array}{c}
\alpha \left( T-T_{c}\right) \Delta ^{2}+\frac{1}{2}\beta _{1}\Delta ^{4}+
\\ 
+\frac{1}{2}\beta _{2}\left( \cos ^{2}\left( 2\chi \right) +\left( \mathbf{%
n\cdot m}\right) ^{2}\sin ^{2}\left( 2\chi \right) \right) \Delta ^{4}%
\end{array}%
\right\} \text{.}  \label{potential}
\end{equation}

This form allows us to make several interesting observations. The crucial
sign is that of $\beta _{2}$. In previous studies\cite{Knigavko,Bel} only $%
\beta _{2}>0$ (so called phase A) was considered. In our case however $\beta
_{2}<0$ and different ground state configurations should be considered. In
phase B the minimization gives, $\mathbf{n=\pm m}$. Note two different
solutions. So that the "vacuum manifold" is 
\begin{equation}
\mathbf{\Delta }=\Delta _{0}\mathbf{n}e^{i\chi }\text{.}  \label{cacuum}
\end{equation}%
Here the range of $\chi $ was enlarged, $-\pi /2<\chi <\pi /2$, to
incorporate $\mathbf{n=\pm m}$. The ground state energy density therefore is
achieved at 
\begin{equation}
\Delta _{0}^{2}=\frac{\alpha \left( T_{c}-T\right) }{\beta _{1}+\beta _{2}}=%
\frac{\alpha \left( T_{c}-T\right) }{\beta }\text{.}  \label{delta0}
\end{equation}%
Mathematically the vacuum manifold in phase B is isomorphic to $S_{2}\otimes
S_{1}/Z_{2}$. This determines the thermodynamics of the superconductor very
much in analogy with the scalar superconductor with $\beta =\beta _{1}+\beta
_{2}$. However the collective modes, the $AC$ conductivity and the magnetic
properties are markedly different \cite{pwave}.

\subsection{Collective excitation modes.}

Here the response of the superconductor in phase B to an external
perturbation, like boundary or magnetic field, is considered. The basic
excitation modes are uncovered by the linear stability analysis very similar
to the so-called Anderson - Higgs mechanism in field theory applied to
(scalar order parameter) superconductivity a long time ago \cite{Weinberg}.
Two basic scales, the coherence length (scale of variations of the order
parameter) and magnetic penetration depth (scale of variations of the
magnetic field),{\huge \ }are obtained from the expansion of the GL energy
to second order in fluctuations around superconducting ground state at zero
field. In the superconducting state the most convenient gauge is the
"unitary" gauge in which the $U\left( 1\right) $ phase of the order
parameter is set to zero, so that we are left with three massive vector
potential fields $A_{1},A_{2},A_{3}$. Chosing the grond state as $\Delta
=\Delta _{0}\left( 0,0,1\right) $, see Fig.2 the order parameter in the
unitary gauge can be parametrized by five real fields

\begin{equation}
\mathbf{\Delta }=\Delta _{0}\left( 1+\varepsilon \right) \left(
R_{1}+iI_{1},R_{2}+iI_{2},1\right) \text{.}  \label{modes}
\end{equation}%
The sponteneous breaking of the space rotation symmetry $O\left( 3\right) $
into its $O\left( 2\right) $ subgroup of rotations within the $x-y$ plane
according to the Goldstone theorem leads to two gapless modes $R_{\alpha }$, 
$\alpha =1,2$. Due to the residual symmetry the Fourier components of the
fluctuation fields can be generally written as combination of the radial and
the tangential components:%
\begin{eqnarray}
R_{\alpha } &=&\left( R_{r}k_{\alpha }+R_{t}\varepsilon _{\alpha \beta
}k_{\beta }\right) /k_{\perp };\text{ }  \label{trdecomposition} \\
\text{\ \ }I_{\alpha } &=&\left( I_{r}k_{\alpha }+I_{t}\varepsilon _{\alpha
\beta }k_{\beta }\right) /k_{\perp } \\
a_{\alpha } &=&\frac{e^{\ast }u_{T}^{1/2}}{\hbar c}\left( A_{r}k_{\alpha
}+A_{t}\varepsilon _{\alpha \beta }k_{\beta }\right) /k_{\perp },  \notag \\
a_{3} &=&\frac{e^{\ast }u_{T}^{1/2}}{\hbar c}A_{3}
\end{eqnarray}%
with $k_{\perp }=\left( k_{x}^{2}+k_{y}^{2}\right) ^{1/2}$. The GL energy to
quadratic order in eight fluctuation fields $\eta =\left\{ \varepsilon
,I_{r},I_{t},R_{r},R_{t},a_{r},a_{t},a_{3}\right\} $,%
\begin{equation}
f=\frac{N\Delta _{0}^{2}}{2}\sum_{\mathbf{k}}\eta _{\mathbf{k}}^{\ast }M\eta
_{\mathbf{k}},  \label{fluctuationmatrix}
\end{equation}%
decomposes into three independent sectors, where $M$ is dimensionless
fluctuation matrix.

1. Three massive fields $I_{r},a_{r}a_{3}$ can mix due to kinetic terms:

\begin{equation}
M_{1}=\left( 
\begin{array}{ccc}
\begin{array}{c}
2\Delta _{0}^{2}\left\vert \beta _{2}\right\vert + \\ 
+u_{T}k^{2}+uk_{\perp }^{2}%
\end{array}
& -i\frac{u}{u_{T}^{1/2}}k_{3} & 0 \\ 
i\frac{u}{u_{T}^{1/2}}k_{3} & 1+\lambda _{T}^{2}k_{3}^{2} & -\lambda
_{T}^{2}k_{3}k_{\perp } \\ 
0 & -\lambda _{T}^{2}k_{3}k_{\perp } & u_{L}+\lambda _{T}^{2}k_{\perp }^{2}%
\end{array}%
\right)  \label{mat1}
\end{equation}%
where $\lambda _{T}^{2}=\hbar ^{2}c^{2}/8\pi u_{T}e^{\ast 2}N\Delta
_{0}^{2}. $

To order $k^{2}$ the eigenvalues are, 
\begin{equation}
E\left( k^{2}\right) /V=N\Delta _{0}^{2}\left( \Omega ^{2}+C_{\perp
}^{2}k_{\perp }^{2}+C_{\parallel }^{2}k_{3}^{2}\right) \text{.}
\label{spectrumgen}
\end{equation}%
The values of the gaps in the spectrum $\Omega ^{2}$ and and corresponding
velocities in directions perpendicular and parallel to the vector order
parameter (taken to be $z$), $C_{\perp },C_{\parallel }$ are given in Table
I\bigskip\ 

2. The tangential massive, $I_{t},a_{t}$, sector

\begin{equation}
M_{2}=\left( 
\begin{array}{cc}
2\left\vert \beta _{2}\right\vert \Delta _{0}^{2}+u_{T}k^{2} & i\frac{u}{%
u_{T}^{1/2}}k_{3} \\ 
-i\frac{u}{u_{T}^{1/2}}k_{3} & 1+\lambda _{T}^{2}k^{2}%
\end{array}%
\right) \text{.}  \label{mat2}
\end{equation}

3. The Goldstone transverse mode $R_{t}$ does not mix, $M_{3}=u_{T}k^{2}$,
so that $E_{6}/V=N\Delta _{0}^{2}u_{T}k^{2}$.

4. Higgs and Goldstone radial collective modes $\varepsilon ,R_{r}$ form a $%
2\times 2$ matrix:

\begin{equation}
M_{4}=\left( 
\begin{array}{cc}
u_{T}k^{2}+uk_{\perp }^{2} & uk_{\perp }k_{3} \\ 
uk_{\perp }k_{3} & u_{T}k^{2}+uk_{3}^{2}+\Delta _{0}^{2}\frac{\beta }{2}%
\end{array}%
\right) \text{.}  \label{mat4}
\end{equation}

\subsection{Coherence length and penetration depth for massive collective
modes}

The six "massive" fields, $\varepsilon ,I_{\alpha }$ and $\mathbf{A}$ with
different longitudinal and transversal characteristic lengths: $l_{\perp
}=C_{\perp }/\Omega ;l_{\parallel }=C_{\parallel }/\Omega $ ,and the same
for $\xi $ and $\lambda $, see Table I.

This is different compared with the one component (singlet) superconductor
in two respects. First the number of Higgs modes is larger since in addition
to the superfluid density determined by $\varepsilon ,$ there are two
additional components $I_{r}$ and $I_{t}$. Second, as mentioned above, since
the superconducting condensate is oriented, one has two different
velocities. As far as (massive) photon modes are concerned, the number of
modes remains the same but the anisotropy persists. Let us start with m

Here definitions of the coherence lengths (of fluctuation of the superfluid
density $\varepsilon $ perpendicular and parallel to the direction of the
order parameter $\mathbf{n}$), correlation lengths of the relative weights
between different components of the vector order parameter ($I_{r}$,$I_{t}$)
and screening lengths (vector potential $\mathbf{A}$ fluctuations):

\begin{eqnarray}
\xi _{\perp }^{2} &=&\frac{2u_{L}}{\beta \Delta _{0}^{2}};\text{ \ }\xi
_{\parallel }^{2}=\frac{u_{T}}{u_{L}}\xi _{\perp }^{2};  \label{coherence} \\
l_{\perp }^{2} &=&\frac{u_{T}}{2\left\vert \beta _{2}\right\vert \Delta
_{0}^{2}};\text{ \ \ }l_{\parallel }^{2}=\frac{u_{L}}{u_{T}}l_{\perp }^{2}; 
\notag \\
\text{\ }\lambda _{T}^{2} &=&\frac{\hbar ^{2}c^{2}}{8\pi u_{T}e^{\ast
2}N\Delta _{0}^{2}},\text{\ }\lambda _{L}^{2}=\frac{u_{T}}{u_{L}}\lambda
_{T}^{2}\text{.}  \notag
\end{eqnarray}%
In the isotropic superconductor one recovers the standard formulas since $%
u_{T}=u_{L}$

Our calculation in the previous Section for the Dirac semi-metal, see Eq.(%
\ref{u_res}), demonstrate that both are quite different since $%
u_{L}/u_{T}=1/32<<1$. This is obviously of great importance for large
magnetic field properties of such superconductors and will be discussed
below.

\section{Magnetic and optical properties}

\subsection{Strong magnetic fields: is there an upper critical field $H_{c2}$%
?}

In strong homogeneous magnetic field $H$ (assumed to be directed along the $%
z $ axis) superconductivity typically (but not always, see an example of the 
$p $-wave superconductor that develops flux phases \cite{pwave}) disappears
at certain critical value $H_{c2}$. This bifurcation point is determined
within the GL framework by the lowest eigenvalue of the linearized GL
equations. This is an exact requirement of stability of the normal phase\cite%
{Ketterson,Alama}. The linearized GL equation Eq.(\ref{GLEZ}) reads:%
\begin{equation}
\left[ \left( a-u_{T}\mathcal{D}^{2}\right) \delta _{ij}-\frac{u}{2}\left\{ 
\mathcal{D}_{i},\mathcal{D}_{j}\right\} \right] \Delta _{j}=0\text{,}
\label{linearized GL}
\end{equation}%
where coefficients are in Eq.(\ref{u_res}), and $u=u_{L}-u_{T}$. We use the
Landau gauge, $A_{x}=H_{c2}y;$ $A_{y}=A_{z}=0$. Assuming translation
symmetry along the field direction, $\partial _{z}\Delta _{i}=0$, the
operators of the eigenvalue problem depend on $x$ and $y$ only.

Since we have three components of the order parameter, there are three
eigenvalues. It is easily seen from Eq.(\ref{linearized GL}) that the $z$-
component of the order parameter $\Delta _{z}$ parallel to the external
field direction is independent of the other two, $\Delta _{x},\Delta _{y}$,
leading to the ordinary Abrikosov value:%
\begin{equation}
-u_{T}\mathcal{D}^{2}\Delta _{z}=-a\Delta _{z}\rightarrow H_{c2}^{\parallel
}=\frac{\Phi _{0}}{2\pi \xi _{T}^{2}}\text{,}  \label{transversalHc2}
\end{equation}%
\ where $\xi _{T}^{2}$ $\equiv u_{T}$ (see Eq.(\ref{u_res}). To avoid
confusion with customary notations for layered materials (like high $T_{c}$
cuprates), the material that is modelled here is isotropic and "parallel",
"perpendicular" and refer to the relative orientation of the magnetic field
to the vector order parameter rather than to a layer. The orientation of the
order parameter in isotropic material considered here, due to degeneracy of
the ground state, is determined by the external magnetic field as we
exemplify next.

The two remaining eigenvalues\ involving only the order parameter components 
$\Delta _{x}$ and $\Delta _{y}$ perpendicular to the field (see Fig.3) are
obtained from diagonalizing the "Hamiltonian":

\begin{eqnarray}
\mathcal{H}\left( 
\begin{array}{c}
\Delta _{x} \\ 
\Delta _{y}%
\end{array}%
\right) &=&-a\left( 
\begin{array}{c}
\Delta _{x} \\ 
\Delta _{y}%
\end{array}%
\right) ;  \label{2DVar} \\
\mathcal{H} &\mathcal{=}&\mathcal{-}\left( 
\begin{array}{cc}
u_{T}\mathcal{D}_{y}^{2}+u_{L}\mathcal{D}_{x}^{2} & \frac{u}{2}\left\{ 
\mathcal{D}_{x},\mathcal{D}_{y}\right\} \\ 
\frac{u}{2}\left\{ \mathcal{D}_{x},\mathcal{D}_{y}\right\} & u_{T}\mathcal{D}%
_{x}^{2}+u_{L}\mathcal{D}_{y}^{2}%
\end{array}%
\right) \text{.}  \notag
\end{eqnarray}

This nontrivial eigenvalue problem fortunately can be solved exactly, see
Appendix C. The lowest eigenstate being a superposition of just two lowest
even Landau levels, $\left\vert 0\right\rangle $ and $\left\vert
2\right\rangle $ are given. The lowest of these eigenvalues is%
\begin{equation}
\frac{e^{\ast }H_{c2}^{\perp }}{\hbar c}\left( 
\begin{array}{c}
\frac{3}{2}\left( u_{T}+u_{L}\right) \\ 
-\sqrt{3\left( u_{T}^{2}+u_{L}^{2}\right) -2u_{T}u_{L}}%
\end{array}%
\right) =\alpha \left( T_{c}-T\right) \text{.}  \label{eigenvalue}
\end{equation}%
The corresponding critical field $H_{c2}^{\perp }$ ("perpendicular" refers
to the order parameter direction (Fig.3) ) that can be expressed via an
effective "perpendicular" coherence length, 
\begin{equation}
H_{c2}^{\perp }=\frac{\Phi _{0}}{2\pi \left( \frac{3}{2}\left( \xi
_{L}^{2}+\xi _{T}^{2}\right) -\sqrt{3\xi _{L}^{4}+3\xi _{T}^{4}-2\xi
_{L}^{2}\xi _{T}^{2}}\right) }\text{.}  \label{Hc2perp}
\end{equation}%
\bigskip

It is always larger than $H_{c2}^{\parallel }$, and therefore is physically
realized. The upper field $H_{c2}^{\perp }$ becomes infinite at $%
r_{c}=u_{L}/u_{T}=\left( 13-4\sqrt{10}\right) /3\simeq 0.117$. This means
that in such material superconductivity persists at any magnetic field like
in some $p$- wave superconductors. It was found in Section IV that for the
simplest Dirac semi-metal, $r=1/32<r_{c}$ see Eq.(\ref{u_res}). Thus there
is no upper critical field in this case. Of course, different microscopic
models that belong to the same universality class, might have higher $r$. In
any case the Abrikosov lattice is expected to be markedly different from the
conventional one and even from the vector order parameter model studied in 
\cite{Knigavko}.

\subsection{Dissipative dynamics}

The set of the GL equations corresponding to this energy are obtained by
variation with respect to $\Delta _{j}^{\ast }$ and $A_{i}$. The first is
the time dependent GL equation (the covariant derivative is replaced by
partial since$\mathcal{\ }$in a superconductor the scalar potential can be
taken to be zero on the mesoscopic scale):

\begin{equation}
-\Gamma \mathcal{\partial }_{t}\Delta _{i}=\frac{\delta F}{\delta \Delta
_{i}^{\ast }}\text{.}  \label{TDGL}
\end{equation}%
This should be supplemented by the Maxwell equation including the normal
metal contribution the the current $\mathbf{J}^{n}=\mathbf{J}-\mathbf{J}^{s}$%
, $J_{i}^{s}=\frac{\ ie^{\ast }N}{\ \hbar \ }\left( u_{T}\Delta _{j}^{\ast }%
\mathcal{D}_{i}\Delta _{j}+u\Delta _{j}^{\ast }\mathcal{D}_{j}\Delta
_{i}\right) +cc$. This determines the dynamics of the vector potential:%
\begin{equation}
\frac{\ \sigma _{n}}{c\ }\partial _{t}A_{i}=J_{i}^{s}-\frac{c}{4\pi }\left(
\partial ^{2}\delta _{ij}-\partial _{i}\partial _{j}\right) A_{j}=J_{i}\text{%
.}  \label{Maxwell}
\end{equation}

In the small fluctuations approximation the dominant role is played by the
two Goldstone modes, $R_{t},R_{r}$, due to spontaneous breaking of the 3D
rotation symmetry. The $R_{t}$ mode is still isotropic, while $R_{r}$ is not
Neglecting the massive excitations the dissipative dynamics of the diffusion
type is governed by%
\begin{eqnarray}
\mathcal{\partial }_{t}R_{t} &=&D_{T}k^{2}R_{t};  \label{diffusion} \\
\mathcal{\partial }_{t}R_{r} &=&\left( D_{T}k_{\perp
}^{2}+D_{L}k_{3}^{2}\right) R_{r}\text{, }  \notag
\end{eqnarray}%
where $D_{T,L}=\frac{N}{\Gamma }u_{T,L}$. The diffusion coefficient of this
equation is anisotropic and is discussed in Section VIII. This would lead to
increase in thermal conductivity inside the superconducting state even at
low temperature.

\subsection{The AC conductivity}

In external AC field represented by (no spatial dispersion), $\mathbf{A}=%
\frac{ic}{\omega }\mathbf{E}\left( \omega \right) e^{i\omega t}$, one
obtains in linear response

\begin{equation}
J_{i}=-\frac{2ie^{\ast 2}N\Delta _{0}^{2}}{\omega \hbar ^{2}}\left(
u_{T}E_{i}+uE_{3}\delta _{3i}\right) +\sigma _{n}E_{i}.  \label{sigma}
\end{equation}%
Therefore the conductivity tensor reads:%
\begin{equation}
\left[ \sigma \left( \omega \right) \right] _{ij}=\left( 
\begin{array}{ccc}
\sigma _{n}-\sigma _{T}^{s} & 0 & 0\  \\ 
0 & \sigma _{n}-\sigma _{T}^{s} & 0 \\ 
0 & 0 & \sigma _{n}-\sigma _{L}^{s}%
\end{array}%
\right) \text{,}  \label{condtensor}
\end{equation}%
where $\sigma _{T,L}^{s}\left( \omega \right) =\frac{2ie^{\ast 2}N\Delta
_{0}^{2}}{\omega \hbar ^{2}}u_{T,L}$. The AC conductivity this is different
for the order parameter and the perpendicular directions. One of the
interesting consequences of this phenomenon is rotation of the polarization
of microwave that passes the DSM film.

\subsection{Rotation of the polarization of the microwave}

The material becomes optically active, i,e. the polarization of the
electromagnetic wave rotates. The dispersion relation is:

\begin{equation}
-\frac{ic^{2}k^{2}}{4\pi \omega }\sigma ^{-1}\left( \omega \right) \mathbf{B}%
=\mathbf{B}\text{.}  \label{Faraday}
\end{equation}%
The perpendicular to the order parameters are eigenvectors with eigenvalue $-%
\frac{ic^{2}k^{2}}{4\pi \omega \left( \sigma _{n}-\sigma _{T}^{s}\right) }$,
\ while the third eigenvector $\left( 0,0,1\right) $ has the eigenvalue is $-%
\frac{ic^{2}k^{2}}{4\pi \omega \left( \sigma _{n}-\sigma _{L}^{s}\right) }$.
Assume that the incident electromagnetic wave described on the surface of
DSM by the magnetic field $\mathbf{B}_{0}$ is perpendicular to the order
parameter direction taken as $z$ (see Fig.4).

\bigskip Without loss of generality it can be taken as $\ x$ (due to the
residual $O\left( 2\right) $ rotation symmetry). The Fourier component of
the magnetic field in the $y-z$ plane are 
\begin{eqnarray}
B_{y} &=&B_{0}\exp \left( ik^{(2)}x\right)  \label{decomposition} \\
B_{z} &=&B_{0}\exp \left( ik^{(3)}x\right) \text{.}  \notag
\end{eqnarray}%
The complex wave vectors are 
\begin{eqnarray}
k^{(2)} &=&\sqrt{\frac{1+\sqrt{\lambda _{T}^{4}/\delta ^{4}+1}}{2}\ }\left( -%
\frac{1}{\lambda _{T}}+i\frac{\lambda _{T}}{\delta ^{2}}\right)  \label{k} \\
k^{\left( 3\right) } &=&\sqrt{\frac{1+\sqrt{\lambda _{L}^{4}/\delta ^{4}+1}}{%
2}}\left( -\frac{1}{\lambda _{L}}+i\frac{\lambda _{L}}{\delta ^{2}}\right) 
\text{.}  \notag
\end{eqnarray}%
Here the screening length is defined as usual $\delta =\left( c/4\pi \omega
\sigma _{n}\right) ^{1/2}$ .

The penetration of the microwave radiation into the sample surface
demonstrates differences for AC component parallel and perpendicular to the
direction of the order parameter resulting in effective rotation of the
incident wave vector like at Faraday effect.

\section{Discussion and conclusions}

\subsection{The vector nature of the order parameter}

The physical properties of the triplet superconductivity appearing in 3D
Dirac semi-metals were considered. Starting from the microscopic model of
the isotropic Dirac semi-metal, the Ginzburg-Landau energy for this field is
derived using the Gor'kov technique. The properties of the triplet
superconductor phase of the Dirac semi-metal has extremely unusual features
that we would like to associate qualitatively with the characteristics of
the Cooper pair. The superconducting state generally is a Bose - Einstein
condensate of composite bosons - Cooper pairs, classically described by the
Ginzburg - Landau energy as a functional of the order parameter. In the
present case the Cooper boson is described by a \textit{vector field }$%
\Delta _{i}\left( \mathbf{r}\right) $. In this respect it is reminiscent to
phonon and vector mesons in particle physics\cite{Weinberg}.

Vector fields generally have both the orbital and internal degrees of
freedom often called polarization. The internal degree of freedom might be
connected to the "valley" degree of freedom of constituents of the composite
boson. We have provided evidence that the Cooper pair in DSM has finite
orbital momentum, albeit, as will be shown shortly, the spin magnetic moment
is zero. Microscopically the unusual nature is related to the presence of
the valley degeneracy in Dirac semi-metal. While in a single band
superconductor the Pauli principle requires a triplet Cooper pair to have
both odd angular momentum and spin, it is no longer the case in the Dirac
semi-metal.

A massive bosonic vector field in isotropic situation (the case considered
here) generally have distinct transversal and longitudinal polarizations
(massless fields like photons in dielectric do not possess the longitudinal
degree of freedom). The results for collective modes in triplet
superconductor in DSM demonstrate pronounced disparity between dispersion of
various polarizations, see Table I. In particular we have found sound
velocity of the two Goldstone modes and, screening lengths of three gapped
photon modes and three coherence lengths of the other (Higgs) gapped modes.
These all have an impact on transport, optical and magnetic properties of
these superconductors.

\subsection{Estimates of the characteristics of the collective modes and
Faraday effect in a typical Dirac semi-metal}

Substituting the values of parameter of the vector GL equation found in
Section IV into formulas for various coherence lengths described in Section
VI, one obtains, 
\begin{eqnarray}
\xi _{0\perp }^{2} &=&\frac{7\zeta \left( 3\right) }{80\pi ^{2}}\frac{%
v_{F}^{2}\hbar ^{2}}{T_{c}^{2}};\text{ \ }\xi _{0z}^{2}=32\xi _{\perp }^{2};
\label{ksai} \\
l_{0\perp }^{2} &=&32\xi _{\perp }^{2};\text{ \ \ }l_{0z}^{2}=\xi _{\perp
}^{2};  \notag \\
\lambda _{0T}^{2} &=&\frac{3\pi }{32}\frac{\hbar ^{3}c^{2}v_{F}}{e^{\ast
2}\mu ^{2}},\text{ \ \ }\lambda _{0L}^{2}=32\lambda _{0T}^{2}\text{,}  \notag
\end{eqnarray}%
where $\xi _{\perp }^{2}\left( T\right) =\xi _{0\perp }^{2}/\left(
1-T/T_{c}\right) $ etc.

For a typical DSM one estimates the Fermi velocity and chemical potential 
\cite{Potemski} $v_{F}=c/200$, $\mu =0.2eV$, and with the expected critical
temperature \cite{pressureBiSe}, \cite{DasSarma14} $T_{c}=5K$, one obtains
for $\xi _{0\perp }=230nm$ and $\lambda _{0T}=220nm$. This has an impact on
the magnetic flux penetration into this kind of superconductors. The value
of Abrikosov parameter $\kappa =\lambda /\xi $ depends on mutual direction
of the DC magnetic field. While $\kappa _{T}\equiv \lambda _{T}/\xi _{\perp
}\simeq 1$, $\kappa _{L}\equiv \lambda _{L}/\xi _{\perp }\simeq 30$.

Let us estimate the characteristics of the two Goldstone modes, arising in
the triplet superconducting state due to spontaneous breaking of the
rotational $O\left( 3\right) $ symmetry. The isotropic transverse mode $%
R_{t} $ defined in Section III is isotropic and its dynamics is described by
Eq.(\ref{diffusion}) with diffusion constant $D_{T}=Nu_{T}/\Gamma $, while
the anisotropic mode $R_{r}$ involves both the transverse and the
longitudinal constant that is different, $D_{L}=Nu_{L}/\Gamma $. To estimate
these, let us exploit the relation \cite{Ketterson} the time constant as $%
\Gamma =\frac{\pi \hbar N}{8T_{c}}$. Thus%
\begin{eqnarray}
D_{T} &=&\frac{2^{5}\cdot 7\zeta \left( 3\right) }{15\pi ^{3}}\frac{%
v_{F}^{2}\hbar }{T_{c}}=2\cdot 10^{4}\frac{cm^{2}}{s},  \label{Dif} \\
D_{L} &=&\frac{7\zeta \left( 3\right) }{15\pi ^{3}}\frac{v_{F}^{2}\hbar }{%
T_{c}}=620\frac{cm^{2}}{s}.
\end{eqnarray}%
These considerations were made for superconductor without significant
pinning - disorder on the mesoscopic scale. Recently the AC response of the
disordered superconductor was utilized to probe Goldstone modes \cite{Cea}.
We have demonstrated that they are abundant in the triplet DSM
superconductor. Therefore it would be interesting to look for effects of the
Goldstone modes including damping resulting in strong sound absorption in
these systems.

We have calculated the AC conductivity of DSM, see Eq.(\ref{Faraday}), and
applied it to describe an intriguing effect of optical activity of DSM.
According to Eq.(\ref{k}), the polarization vector of the incident beam
rotates while passing a film of thickness $d$ by (see Fig.4) 
\begin{equation}
\tan \phi =\left\vert B_{z}\left( d\right) /B_{y}\left( d\right) \right\vert
=\exp \left[ \left( \lambda _{T}^{-1}-\lambda _{L}^{-1}\right) d\right] ,
\label{tang}
\end{equation}%
under the assumption the skin depth $\delta =\left( c/4\pi \omega \sigma
_{n}\right) ^{1/2}\ $is much larger than both $\lambda _{T}$ and $\lambda
_{L}$. In particular for $\lambda _{L}=32\lambda _{T},\lambda _{T}=220nm,$%
and $d=1mm,$ $\phi \rightarrow \pi /2$ significantly deviation from the
initial $\pi /4$ value. Let us stress that the effect is due to the
difference between the two penetration depths.

\subsection{DSM superconductor under constant magnetic field}

Several new features appear when an external field is applied. The Ginzburg
- Landau model was used to determine upper magnetic field $H_{c2}$. It turns
out that the lowest energy solution is when the order parameter of the
texture orients itself perpendicular to the field direction. We have shown
that the upper field becomes infinite for $u_{L}/u_{T}<\left( 13-4\sqrt{10}%
\right) /3\simeq 0.117$. In particular it is obeyed within our microscopic
model. This means that in such material superconductivity persists at any
magnetic field like in some $p$- wave superconductors. The expression for
the $H_{c2}$ for $u_{L}/u_{T}>0.117$ is given by Eq.(\ref{Hc2perp}).

The vortex physics of strongly type II triplet superconductors of this type
is very rich and some of it has already been investigated in connection with
heavy fermion and other superconductors suspected to possess $p$-wave
pairing. In particular, their magnetic vortices appear as either vector
vortices or so-called skyrmions\cite{Knigavko} - coreless topologically
nontrivial textures. The magnetic properties like the magnetization are very
peculiar and even without a magnetic field the system forms a "spontaneous
flux state". The material therefore can be called a "ferromagnetic
superconductor". The superconducting state develops weak ferromagnetism and
a system of alternating magnetic domains\cite{Bel}. It was noted\cite%
{Knigavko} that the phase is reminiscent to the phase B of superfluid $%
He_{3} $. \cite{Ketterson} (with an obvious distinction that the order
parameter in the later case is neutral rather than charged and tensorial
rather than vectorial).

Since the prediction of the FFLO effect\cite{Ketterson} in low $T_{c}$
superconductors it is well known that at very high magnetic fields the
direct spin - magnetic field coupling on the microscopic level might not be
negligible. The singlet channel Cooper pair is effectively "broken" by the
splitting since the spins of the two electrons are opposite (Pauli
paramagnetic limit). It is not clear what impact it has on Dirac
semi-metals. If the impact is large it could be incorporated as an
additional paramagnetic term in the GL energy. In an isotropic Dirac
superconductor one has only one possible term in the GL energy term linear
in paramagnetic coupling and consistent with symmetries:%
\begin{equation}
F_{par}=N\mu _{p}\int_{\mathbf{r}}i\left( \mathbf{\Delta }^{\ast }\mathbf{%
\times \Delta }\right) \cdot \mathbf{B}\text{,}  \label{FZ}
\end{equation}%
where $\mu _{p}$ is the effective "spin" of the Cooper pair sometimes called
"Zeeman coupling"\cite{Knigavko,Alama}. The single particle Hamiltonian in
magnetic field has the Pauli term $\mu _{B}\mathbf{\Sigma }\cdot \mathbf{B},$%
where the Bohr magneton, $\mu _{B}=e\hbar /2mc$, determines the strength of
the coupling of the spin to magnetic field, with $m$ being the free electron
mass. The direct calculation, see Appendix A, shows that $\mu _{p}=0$.

Acknowledgements. We are indebted to D. Li and M. Lewkowicz for valuable
discussions. Work of B.R. was supported by NSC of R.O.C. Grants No.
98-2112-M-009-014-MY3 and MOE ATU program.

\newpage

\section{Appendix A. Local terms in GL free energy}

\subsection{Critical temperature calculation}

Starting from equation Eq.(\ref{Tc}) the angle integrations result in (for $%
\mu >>T_{D},T_{c})$

\begin{eqnarray}
\frac{1}{g} &=&T\sum\limits_{np}\frac{\mu ^{2}+\omega _{n}^{2}}{\omega
_{n}^{4}+\left( v_{F}^{2}p^{2}-\mu ^{2}\right) ^{2}+2\omega _{n}^{2}\left(
v_{F}^{2}p^{2}+\mu ^{2}\right) }  \label{A1} \\
&=&\frac{\mu ^{2}}{12\pi ^{2}\hbar ^{3}v_{F}^{3}}\int_{\varepsilon
=-T_{D}}^{T_{D}}\frac{\tanh \left( \varepsilon /2T\right) }{\varepsilon }%
\approx \frac{\mu ^{2}}{6\pi ^{2}\hbar ^{3}v_{F}^{3}}\log \frac{2T_{D}\gamma
_{E}}{\pi T_{c}}.  \notag
\end{eqnarray}%
where $\varepsilon =v_{F}p-\mu $. See the last (BCS) integral in\cite{AGD}.

\subsection{ Cubic terms coefficients calculation}

To fix the two coefficients, $\beta _{1}$ and $\beta _{2}$ in Eq.(\ref{GLEZ}%
) we use only two components. The particular case $j=k=l=1$ (the coefficient
of $\psi _{1}^{\ast 2}\psi _{1}$) gives after angle integration

\begin{equation}
N\left( \beta _{1}+\beta _{2}\right) =\frac{2T}{15\pi ^{2}\hbar ^{3}}%
\sum\limits_{n}\int_{p=0}^{\infty }S\left( p,n\right)  \tag{A2}
\end{equation}%
where 
\begin{equation*}
S\left( p,n\right) =\frac{v_{F}^{2}p^{2}\left(
v_{F}^{4}p^{4}+10v_{F}^{2}p^{2}\left( \omega _{n}^{2}-5\mu ^{2}\right)
-15\left( \mu ^{2}+\omega _{n}^{2}\right) ^{2}\right) }{\left(
v_{F}^{4}p^{4}+2v_{F}^{2}p^{2}\left( \omega _{n}^{2}-\mu ^{2}\right) +\left(
\mu ^{2}+\omega _{n}^{2}\right) ^{2}\right) ^{2}}
\end{equation*}

\bigskip

Performing finite integration (the upper bound on momentum, $\mu +T_{D}$,
can be replaced by infinity), one obtains 
\begin{equation}
N\left( \beta _{1}+\beta _{2}\right) =\frac{8\mu ^{2}}{15\pi
^{4}T^{2}v_{F}^{3}\hbar ^{3}}s_{3}\text{,}  \label{A3}
\end{equation}%
where the sum is

\begin{equation}
s_{3}=\sum_{n=0}\frac{1}{\left( 2n+1\right) ^{3}}=\frac{7\zeta \left(
3\right) }{4}\text{.}  \label{A4}
\end{equation}%
Similarly taking $j=l=2,k=1$ (the coefficient of $\Delta _{2}^{\ast 2}\Delta
_{1}$) gives after the angle integration 
\begin{eqnarray}
N\beta _{2} &=&\frac{2T}{15\pi ^{2}\hbar ^{3}}\sum\limits_{n}\int_{p=0}^{%
\infty }  \label{A5} \\
&&\frac{v_{F}^{2}p^{2}\left( 7v_{F}^{4}p^{4}+10v_{F}^{2}p^{2}\left( \omega
_{n}^{2}+3\mu ^{2}\right) +15\left( \mu ^{2}+\omega _{n}^{2}\right)
^{2}\right) }{\left( v_{F}^{4}p^{4}+2v_{F}^{2}p^{2}\left( \omega
_{n}^{2}-\mu ^{2}\right) +\left( \mu ^{2}+\omega _{n}^{2}\right) ^{2}\right)
^{2}}  \notag \\
&=&-\frac{4\mu ^{2}}{15\pi ^{4}v_{F}^{3}\hbar ^{3}T^{2}}s_{3}\text{.}  \notag
\end{eqnarray}%
resulting in Eq.(\ref{betas}).

\subsection{Effect of the Pauli interaction term}

The single particle Hamiltonian in magnetic field is 
\begin{equation}
\text{\ }\widehat{K}=-iv_{F}\hbar \mathcal{D}\mathbf{\cdot \alpha }-\mu +\mu
_{B}\mathbf{\Sigma }\cdot \mathbf{B}\text{,}  \label{KZeeman}
\end{equation}%
In order to fix the coefficient of the paramagnetic term linear in both the
order parameter and Pauli coupling it is enough to expand the linearized
Gorkov equations Eq.(\ref{linear1}) to the first order in the spin density.
Normal Greens functions have the following corrections: 
\begin{eqnarray}
D_{Z} &\approx &D-\mu _{B}D\left( \Sigma \cdot B\right) D\text{;}  \label{A6}
\\
\widetilde{D}_{Z} &\approx &\widetilde{D}+\mu _{B}\widetilde{D}\left( \Sigma
^{t}\cdot B\right) \widetilde{D}\text{.}  \notag
\end{eqnarray}%
The Pauli term in Gor'kov equation (after multiplying by $T_{i}^{t}$ and
taking the trace as usual), Eq.(\ref{GLEZ}), therefore is obtained from
expansion of Eq.(\ref{linear1}), 
\begin{eqnarray}
&&\sum_{\omega \mathbf{p}}\text{Tr}\left\{ T_{i}\widetilde{D}\Delta ^{\ast
}D\right\}  \label{A7} \\
&=&-i\mu _{Z}\varepsilon _{ijk}\Delta _{j}^{\ast }B_{k}N=\mu _{B}\mathcal{B}%
_{ijk}^{Z}\Delta _{j}^{\ast }B_{k},  \notag \\
\mathcal{B}_{ijk}^{Z} &=&\sum_{\omega \mathbf{p}}\text{Tr}\left\{ T_{i}^{{}}%
\widetilde{D}\left( \Sigma _{k}^{t}\widetilde{D}T_{j}^{\ast }-T_{j}^{\ast
}D\Sigma _{k}\right) D\right\} \text{,}  \notag
\end{eqnarray}%
The bubble sum is directly evaluated and vanishes $B_{ijk}^{Z}=0$.

\section{Appendix B. Calculation of gradient terms in the GL}

Rotational invariance allows to represent the sum in Eq (\ref{nonlocal _
terms}) terms of coefficients $u_{T}$ and $u_{L}$:%
\begin{eqnarray}
&&-N\left( u_{T}\left( P^{2}\delta _{mj}-P_{m}P_{j}\right)
+u_{L}P_{m}P_{j}\right)  \label{B1} \\
&=&P_{k}P_{l}\sum_{\omega \mathbf{q}}Tr\left\{ T_{m}^{t}\widetilde{D}\left(
\omega ,\mathbf{q}\right) T_{j}^{t\ast }D_{kl}^{\prime \prime }\left( \omega
,\mathbf{q}\right) \right\} ,  \notag
\end{eqnarray}%
where 
\begin{eqnarray}
D_{ij}^{\prime \prime } &=&\frac{2}{\left( q^{2}-\left( i\omega +\mu \right)
^{2}\right) ^{2}}  \label{B2} \\
&&\left\{ q_{j}\alpha _{i}+q_{i}\alpha _{j}+\delta _{ij}\left( i\omega +\mu +%
\mathbf{\alpha \cdot q}\right) +2q_{i}q_{j}D\right\} \text{.}  \notag
\end{eqnarray}%
In particular%
\begin{eqnarray}
Nu_{L} &=&-\sum_{\omega \mathbf{p}}Tr\left\{ T_{z}\widetilde{D}\left( \omega
,\mathbf{p}\right) T_{z}^{\ast }D_{zz}^{\prime \prime }\left( \omega ,%
\mathbf{p}\right) \right\}  \label{B3} \\
&=&\frac{\mu ^{2}}{15\pi ^{4}T^{2}v_{F}\hbar }s_{3}  \notag \\
&=&\frac{7\zeta \left( 3\right) }{60\pi ^{4}}\frac{\mu ^{2}}{T^{2}v_{F}\hbar 
}=\frac{7\zeta \left( 3\right) \left( v_{F}\hbar \right) ^{2}N}{120\pi
^{2}T^{2}}  \notag
\end{eqnarray}%
and%
\begin{equation}
Nu_{T}=-\sum_{\omega \mathbf{p}}Tr\left\{ T_{x}\widetilde{D}\left( \omega ,%
\mathbf{p}\right) T_{x}^{\ast }D_{zz}^{\prime \prime }\left( \omega ,\mathbf{%
p}\right) \right\} =32u_{L}\text{.}  \label{B4}
\end{equation}

\section{Appendix C. Collective modes}

Here details of the calculation of free energy in harmonic approximation are
given. The expansion of order parameter Eq.(\ref{modes}) to quadratic order
is%
\begin{equation}
\mathbf{\Delta /}\Delta _{0}=\left( 1+\varepsilon \right) \left( \delta
_{1},\delta _{2},1\right) =\left( 0,0,1\right) +\left( \delta _{1},\delta
_{2},\varepsilon \right) +\left( \delta _{1}\varepsilon ,\delta
_{2}\varepsilon ,0\right) \text{.}  \label{c1}
\end{equation}%
The gradient terms that do not involve the vector potential $\mathbf{A}$ are:

\begin{eqnarray}
\frac{F_{grad}^{\left( 1\right) }}{\Delta _{0}^{2}} &=&u_{T}\left( \partial
_{j}R_{\alpha }\partial _{j}R_{\alpha }+\partial _{j}I_{\alpha }\partial
_{j}I_{\alpha }+\left( \partial _{j}\varepsilon \right) ^{2}\right)
\label{c2} \\
&&+u\left\{ 
\begin{array}{c}
\partial _{1}R_{1}\partial _{1}R_{1}+\partial _{1}I_{1}\partial _{1}I_{1}+
\\ 
+\partial _{2}R_{2}\partial _{2}R_{2}+\partial _{2}I_{2}\partial
_{2}I_{2}+\left( \partial _{3}\varepsilon \right) ^{2} \\ 
+2\left( 
\begin{array}{c}
\left( \partial _{2}R_{1}\right) \left( \partial _{1}R_{2}\right) + \\ 
+\left( \partial _{2}I_{1}\right) \left( \partial _{1}I_{2}\right) +\left(
\partial _{\alpha }\varepsilon \right) \left( \partial _{3}R_{\alpha }\right)%
\end{array}%
\right)%
\end{array}%
\right\}  \notag
\end{eqnarray}%
The terms involving $\mathbf{A}$ read%
\begin{equation}
\frac{F_{grad}^{\left( 2\right) }}{\Delta _{0}^{2}}=2\frac{e^{\ast }u}{\hbar
c}\left( \partial _{3}I_{\alpha }\right) A_{\alpha }+\left( \frac{e^{\ast }}{%
\hbar c}\right) ^{2}\left( u_{T}A^{2}+uA_{3}^{2}\right)  \label{c3}
\end{equation}

Potential terms result in

\begin{equation}
\frac{F_{pot}}{\Delta _{0}^{2}}=\Delta _{0}^{2}\left\{ \frac{\beta
_{1}+\beta _{2}}{2}\varepsilon ^{2}-2\beta _{2}\left(
I_{1}^{2}+I_{2}^{2}\right) \right\} ,  \label{c4}
\end{equation}%
while the magnetic energy is:%
\begin{equation}
\frac{F_{mag}}{\Delta _{0}^{2}}=\frac{\hbar ^{2}c^{2}}{8\pi e^{\ast
2}N\Delta _{0}^{2}}A_{i}^{\ast }\left( k^{2}\delta _{ij}-k_{i}k_{j}\right)
A_{j}  \label{c5}
\end{equation}%
The fluctuation matrix $M$ in Eq.(\ref{fluctuationmatrix}) thus was
constructed.

\section{Appendix D. Exact solution for upper critical magnetic field}

In this Appendix the matrix $\mathcal{H}$ defined in Eq.(\ref{2DVar})
determining the perpendicular upper critical field is diagonalized
variationally.

\subsection{Creation and annihilation operators}

Using Landau creation and annihilation operators in units of magnetic length 
$\frac{e^{\ast }B}{c}=l^{-2}$ for the state with $k_{x}=0$ (independent of $%
x $), so that covariant derivatives are 
\begin{eqnarray}
D_{x} &=&\partial _{x}+iy=iy=\frac{i}{\sqrt{2}}\left( a+a^{+}\right) ;
\label{d1} \\
D_{y} &=&\partial _{y}=\frac{1}{\sqrt{2}}\left( a-a^{+}\right) \text{.} 
\notag
\end{eqnarray}%
In terms of these operators the matrix operator $\mathcal{H}$ takes a form:%
\begin{eqnarray}
\mathcal{H} &=&u_{T}+\frac{u}{2}+\mathcal{V};  \label{d2} \\
\mathcal{V}_{11} &=&2u_{T}a^{+}a+\frac{u}{2}\left(
a^{2}+a^{+2}+2a^{+}a\right) ;  \notag \\
\mathcal{V}_{12} &=&\mathcal{V}_{21}=\frac{iu}{2}\left( a^{+2}-a^{2}\right) ;
\notag \\
\mathcal{V}_{22} &=&2u_{T}a^{+}a-\frac{u}{2}\left(
a^{2}+a^{+2}-2a^{+}a\right) \text{.}  \notag
\end{eqnarray}

The exact lowest eigenvalue is a combination of two lowest Landau levels.
Indeed applying the operator $\mathcal{V}$ on a general vector on the
subspace gives 
\begin{eqnarray}
&&\mathcal{V}\left( 
\begin{array}{c}
\alpha \left\vert 0\right\rangle +\beta \left\vert 2\right\rangle \\ 
\gamma \left\vert 0\right\rangle +\delta \left\vert 2\right\rangle%
\end{array}%
\right)  \label{d3} \\
&=&\left( 
\begin{array}{c}
\frac{u}{\sqrt{2}}\left( i\delta -\beta \right) \left\vert 0\right\rangle \\ 
-\left( u\left( \frac{\alpha }{\sqrt{2}}+\frac{i\gamma }{\sqrt{2}}+2\beta
\right) +4u_{T}\beta \right) \left\vert 2\right\rangle \\ 
+\frac{u}{2}\left( -\beta -i\delta \right) \left\vert 4\right\rangle \\ 
+\frac{u}{\sqrt{2}}\left( i\beta +\delta \right) \left\vert 0\right\rangle
\\ 
+\left( u\left( -\frac{i\alpha }{\sqrt{2}}+\frac{\gamma }{\sqrt{2}}-2\delta
\right) -4u_{T}\delta \right) \left\vert 2\right\rangle \\ 
+\frac{u}{2}\left( -i\beta +\delta \right) \left\vert 4\right\rangle%
\end{array}%
\right) \text{.}  \notag
\end{eqnarray}%
For $\delta =i\beta $, higher Landau levels decuple and one gets eigenvalue
equations%
\begin{equation}
\left\vert 
\begin{array}{ccc}
-v & -\sqrt{2}u & 0 \\ 
-\frac{u}{\sqrt{2}} & -4u_{T}-2u-v & -\frac{iu}{\sqrt{2}} \\ 
0 & ui\sqrt{2} & -v%
\end{array}%
\right\vert =0\text{,}  \label{d4}
\end{equation}%
resulting in three eigenvalues of $\mathcal{H}$ 
\begin{eqnarray}
h^{\left( 1\right) } &=&u_{T}+u/2,h^{\left( \pm \right) }  \label{d5} \\
&=&3u_{T}+\frac{3}{2}u\pm \sqrt{4u_{T}^{2}+4u_{T}u+3u^{2}}\text{.}  \notag
\end{eqnarray}

Fig. 1. A complex vector field can be written as $\mathbf{\Delta }=\Delta
\left( \mathbf{n}\cos \chi \mathbf{+}i\mathbf{m}\sin \chi \right) $, where $%
\mathbf{n}$ and $\mathbf{m}$ are arbitrary unit vectors and $0<\chi <\pi /2$.

\bigskip

Fig.2. \ Fluctuations of the order parameter in the unitary gauge can be
parametrized by the five real fields $\mathbf{\Delta }=\Delta _{0}\left(
1+\varepsilon \right) \left( R_{1}+iI_{1},R_{2}+iI_{2},1\right) .$

Fig.3.

Superconductivity arises from the normal state when the order parameter is
formed in direction perpendicular to the magnetic field.

Fig. 4

The polarization vector of the incident beam rotates while passing a film of
thickness $d$ by angle $\phi $ $\tan \phi =\left\vert B_{z}\left( d\right)
/B_{y}\left( d\right) \right\vert =\exp \left[ \left( \lambda
_{T}^{-1}-\lambda _{L}^{-1}\right) d\right] $.

\bigskip

\newpage

\end{document}